\documentclass[12pt,a4paper]{article}
 \usepackage{subfigure}
\usepackage{graphicx}
\usepackage{wrapfig}

\begin{document}
\textwidth=135mm
 \textheight=200mm
 \parindent=0em
 
\begin{center}
{\bfseries Study of the hadronisation process from single hadron and hadron-pair production in SIDIS at COMPASS}

\vskip 5mm
N. Makke$^{\dag}$ \textit{\small{on behalf of the COMPASS Collaboration}}
\vskip 5mm
{\small {\it $^\dag$ INFN/University of Trieste, via A. Valerio 2, 34127 Trieste,  Italy}}
\\
\end{center}

\centerline{\bf Abstract}
Recent results on single-hadron and hadron-pair multiplicities measurement in semi-inclusive deep inelastic muon scattering on nuclear target at the COMPASS experiment at CERN are presented and discussed. 

\section{\label{sec:intro}Introduction}
The hadronisation process describes the collinear transition of a patron into a final-state hadron carrying a fractional energy $z$. It is relevant in every hard scattering reaction where hadrons are created and observed and the final state. Nowadays the hadronisation process remains at a very preliminary stage of knowledge with a growing interest in more accurate and precise measurements of fragmentation functions. Two aspects of this process are studied and discussed: Single-hadron (FFs) and hadron pair (DFFs) fragmentation functions. While the FFs represent a key element in the flavor separation of polarized Parton Distribution Functions (PDFs) and play a particular role in the strange quark polarization ($\Delta S$) puzzle, the DFFs are crucial to access the transversity function, a poorly known cornerstone in the spin structure of the nucleon.  Currently, while pion FFs are known with limited precision and kaon FFs are poorly known, the  situation for the DFFs turns out to be behind the schedule and nor experimental neither theoretical studies have been performed yet.

\begin{wrapfigure}{r}{0.3\textwidth}
\vspace{-31pt}
\begin{center}
\includegraphics[width=0.2\textwidth]{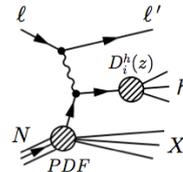}
\end{center}
\vspace{-20pt}
\caption{Schematic representation of the SIDIS reaction.}
\label{SIDIS}
\vspace{-10pt}
\end{wrapfigure}

Semi-inclusive DIS (SIDIS) (see Fig.\ref{SIDIS}), where a final-state hadron is detected in coincidence with the scattered lepton, is a tool to study the hadron creation mechanism. It allows flavor and charge separation, disentangles between  quark and anti-quark hadronisation and provides a wide range of the hard 

energy scale $Q^{2}$. The SIDIS cross section, however,  depends upon the PDFs which are well constrained except for the poorly known strange quark distribution ($s(x)$) which still carries large uncertainties. The most relevant observable to measure in SIDIS is the hadron multiplicity defined in Eq. \ref{MulDef} for the single-hadron case. It depends simultaneously upon the PDFs ($q(x,Q^{2}$)) and the FFs ($D_{q}^{h}(z,Q^{2})$).

\begin{equation}
M^{h}(x,z,Q^{2}) = \frac{d\sigma^{h}/dxdzdQ^{2}}{\sigma^{DIS}(x,Q^{2})} = \frac{\sum_{q}e_{q}^{2}q(x,Q^{2}) \cdot D_{q}^{h}(z,Q^{2})}{\sum_{q} e_{q}^{2}q(x,Q^{2})}
\label{MulDef}
\end{equation}

Experimentally, the hadron multiplicity defines the averaged number of hadrons produced per DIS event, normalized by the acceptance correction factor which takes into account the limited angular and geometrical acceptance coverage by the experimental setup as well as the kinematic smearing.

\section{\label{sec:multiplicities}Single-hadron: Data Analysis \& Results}
\begin{figure}[h!]
\centering
\subfigure[$M^{\pi^{\pm}/K^{\pm}}(x,z)$     ]{\label{xz}\includegraphics[height=6.8cm,width=0.5\textwidth]{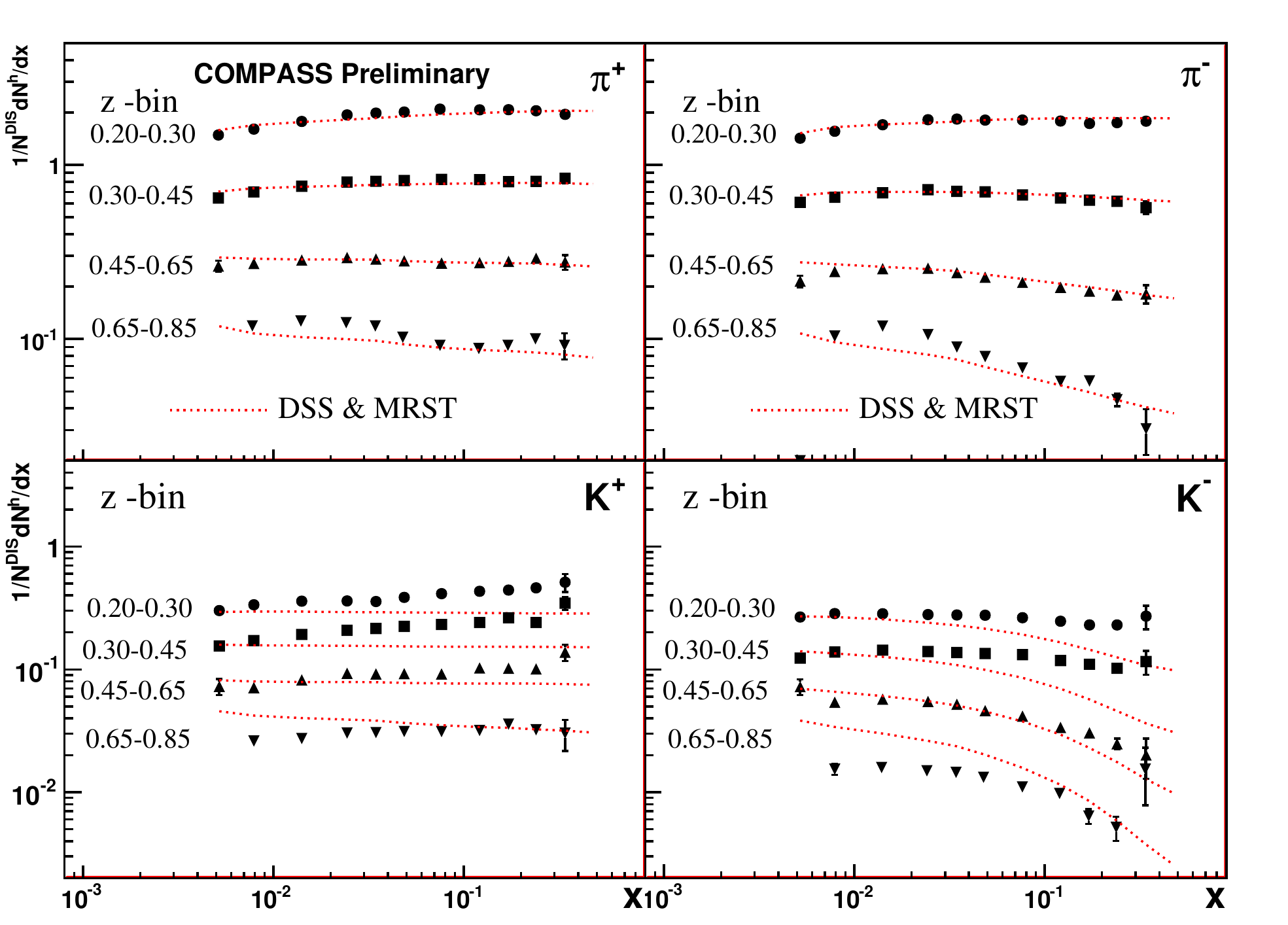}}
\subfigure[$M^{\pi^{\pm}/K^{\pm}}(Q^2,z)$]{\label{qz}\includegraphics[height=6.8cm,width=0.5\textwidth]{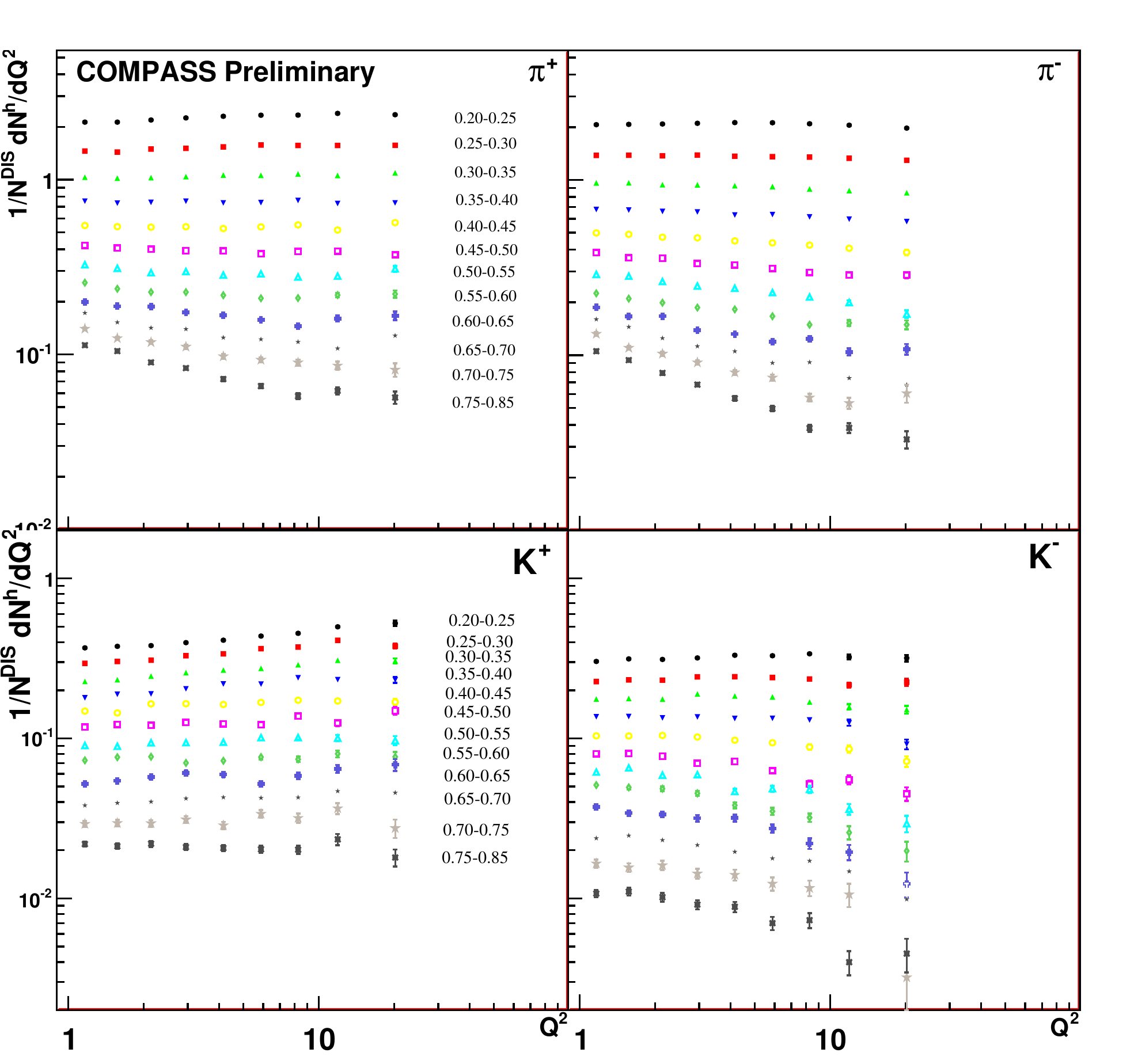}}
\caption{Pion (upper row) and kaon (lower row) multiplicities versus ($x$,$z$) (left) and ($Q^{2}$,$z$) (right). Only statistical errors are shown.}
\label{mulpik}
\end{figure}

The data were collected at the COMPASS experiment at CERN using a 160 GeV/C muon beam and a deuteron target ($^{6}$LiD) during the 2004 data taking. The COMPASS apparatus consists of two stages, each one is equipped with a magnet and a set of tracking detectors. COMPASS offers the particle identification ability with the RICH detector which restricts the momentum coverage to [$3/10,50$] GeV/c for $\pi$/K. The DIS sample is selected using cuts on the photon virtuality $Q^{2} > 1$ (GeV/c)$^{2}$, on the lepton energy fraction transferred to the virtual photon $0.1 < y < 0.9$. To avoid kinematic regions with large and non-uniform acceptance, a cut on the invariant mass of the hadronic system is applied $W > 7$ GeV/c. The energy fraction of the photon transferred to the hadron is restricted to the range [$0.2,0.85$] to avoid any contamination from the target fragmentation region and exclusive reactions.\\ 
Figure \ref{xz} shows the resulting $\pi$ and K multiplicities versus $x$ and $z$ compared to LO theoretical predictions calculated using DSS FFs and MRST04 PDFs. Experimental data points show a fair agreement with predictions for pions except for $z>0.65$ while for kaons, significant discrepancies are observed in the full $z$ range highlighting that COMPASS data favor different FFs than DSS \cite{LSSFF}. The $Q^{2}$ dependence of the hadron multiplicities is also studied with a fine binning in $z$, as shown in Fig. \ref{qz}. A non-negligible $Q^{2}$ dependence is observed for pions and kaons.
Systematic uncertainties were estimated and found to be $\sim 5(10)\%$ for $\pi$(K), where the largest contribution is due to particle identification (more details can be found in \cite{MakkeThesis}). In order to reduce the systematic errors, the data set collected in 2006 on deuteron target is being analyzed.

A recent preliminary NLO QCD fit of fragmentation functions to these results was performed by E. Leader et \textit{al}. \cite{LSSFF}. The results are found to be comparable with DSS FFs only for pions, significant disagreement is found for kaons. The COMPASS results showed an important contribution from gluons.

\section{Hadron pair: Data Analysis \& Results}

The transversity function remains a poorly known element in the nucleon spin structure and is mainly addressed by the collins effect in SIDIS transverse spin asymmetries. An alternative and complementary approach is based on a spin asymmetry in the semi-inclusive $\textit{l}N^{\uparrow} \rightarrow \textit{l'}(h_{1}h_{2})X$ process where two unpolarized hadrons emerge from the hadronisation of the struck quark in the final state. The kinematic remains similar to single-hadron case except for hadronic variables, where a hadron pair carries a fractional energy $z = z_{1} + z_{2}$, with a total momentum $P=P_{1}+P_{2}$ and an invariant mass $M_{inv}^{2}$ which naturally represents a second scale for the fragmentation. In this case, the

\begin{wrapfigure}{r}{0.35\textwidth}
\vspace{-25pt}
\begin{center}
\includegraphics[height=3.cm,width=0.2\textwidth]{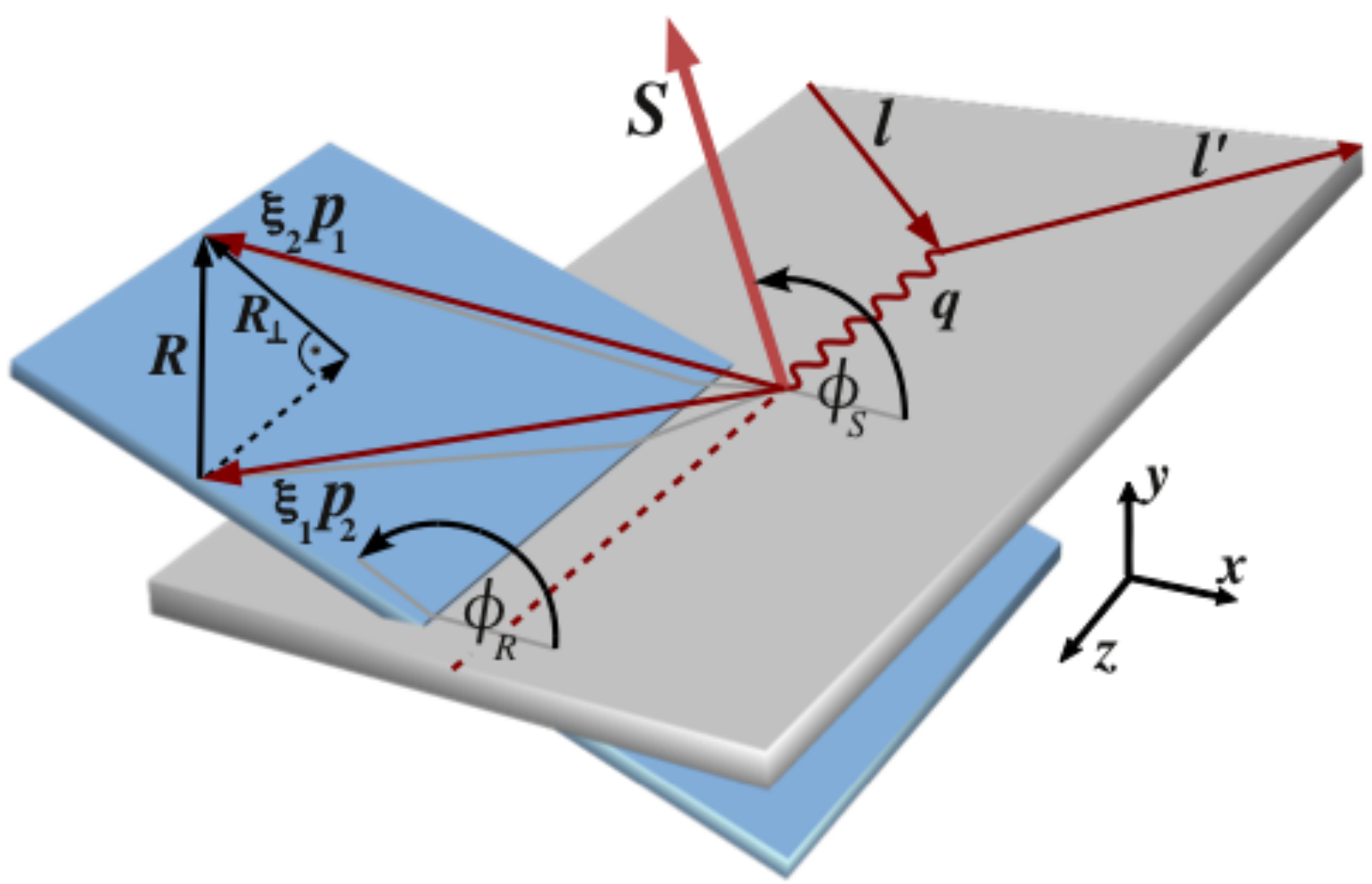}
\end{center}
\vspace{-20pt}
\caption{The hadron pair production in SIDIS.}
\label{SIDIS2h}
\vspace{-10pt}
\end{wrapfigure}
\begin{figure}[h!]
\centering
\vspace{-3mm}
\includegraphics[height=5cm,width=0.7\textwidth]{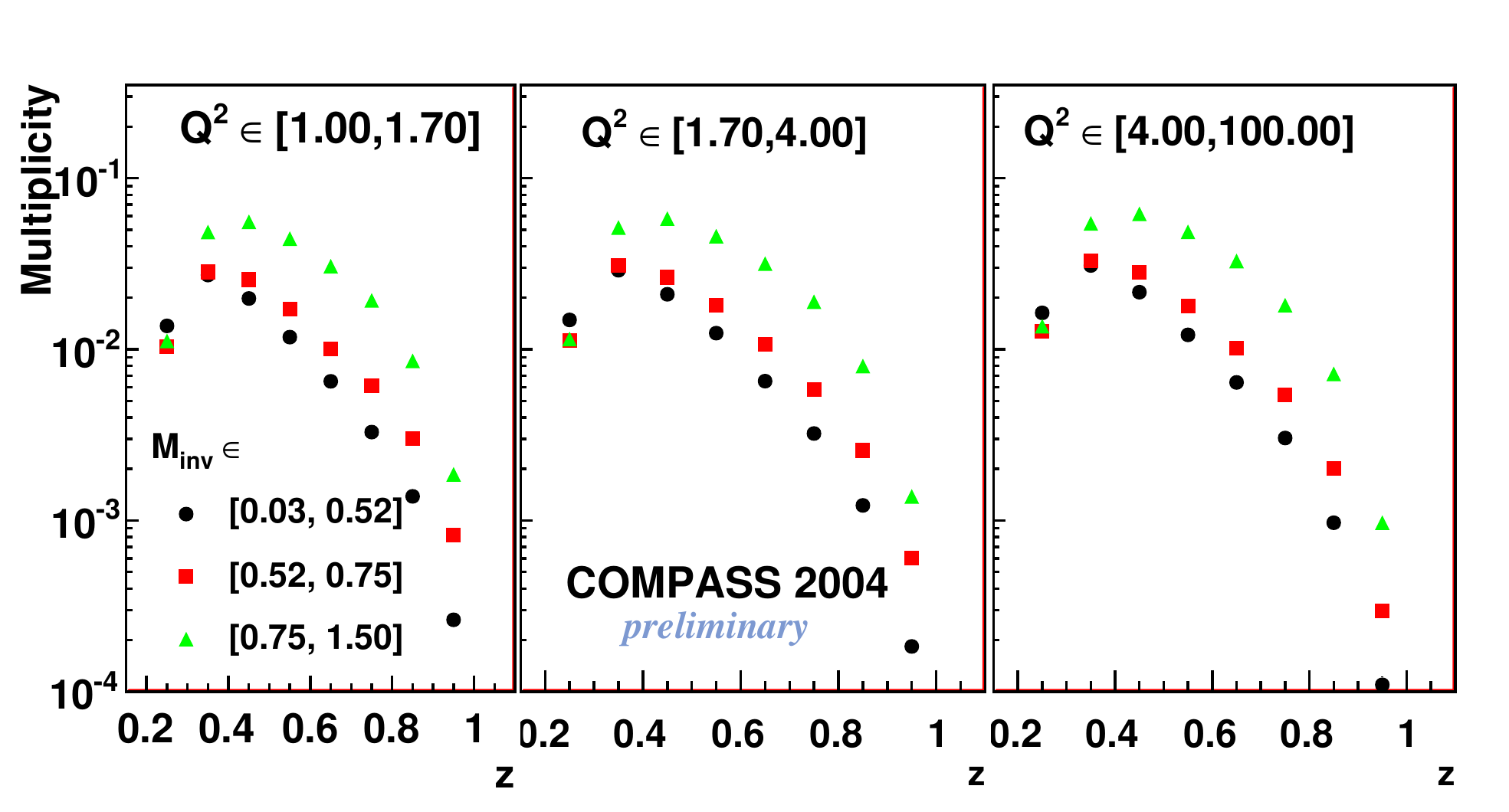}
\vspace{-4mm}
\caption{$h^{+}h^{-}$ multiplicities measured versus $z$, $M_{inv}$ and $Q^{2}$.}
\label{pairs}
\end{figure}


transverse asymmetry is expressed in terms of polarized DFFs ($H_{1,q}^{<h_{1}h_{2}} (z,M_{inv},Q^{2})$) which represents the interference between the fragmentation amplitudes into hadron pairs in relative s and p waves, and in terms of unpolarised DFFs ($D_{q}^{h_{1}h_{2}}(z,M_{inv},Q^{2})$). While the polarized DFFs were recently measured at the BELLE experiment, the unpolarised DFFs are not yet measured. For this goal, we present the first measurement of hadron pair multiplicities in SIDIS at COMPASS using 2004 data collected on a deuteron target. The same kinematic is covered in this measurement except for the energy of the hadronic system, $W > 5$ GeV/c in this case. In addition, each hadron is required to have its fractional energy larger than $0.1$ and its Feynman variable larger than $0.1$ to avoid the target fragmentation region. 
 The resulting hadron pair multiplicities are shown in Fig.\ref{pairs} versus the fractional energy of the pair $z$, the invariant mass of the pair ($M_{inv}$) and $Q^{2}$. While the $Q^{2}$ dependence is found to be negligible up to $z=0.8$, a strongest dependence on $z$ and $M_{inv}$ is observed. No theoretical predictions exist yet for comparison. Systematic errors are estimated to $7\%$.

\section{Conclusions}
Single-hadron and hadron-pair multiplicities have been measured at the COMPASS experiment at CERN using a $160$ GeV/c muon beam and a deuteron target. The results on single-hadron multiplicities favor different set of FFs than currently existing ones, mainly for kaon, and highlight an important contribution of the gluon FFs. The hadron pair multiplicities represent the first measurement in the purpose of extracting DFFs and the transversity.


\end{document}